\documentclass{aastex702}
\usepackage{amsmath,amssymb,booktabs,placeins}
\newcommand{\FiDeR}{\textsc{FiDeR}}
\newcommand{\RL}{\mathcal{D}}
\newcommand{\diag}{\operatorname{diag}}
\shorttitle{FiDeR image denoising}
\shortauthors{Wang et al.}
\journalinfo{arXiv preprint}
\begin{document}
\renewcommand*{\theHequation}{\thesection.\arabic{equation}}
\title{FiDeR: Denoising Astronomical Images at the Observational Resolution}
\correspondingauthor{Lei Wang}
\author[0000-0002-3779-9069]{Lei Wang}
\affiliation{Purple Mountain Observatory, Chinese Academy of Sciences, 10 Yuanhua Road, Qixia District,
Nanjing 210023, China}
\email{leiwang@pmo.ac.cn}
\author[]{Bin Li}
\affiliation{Purple Mountain Observatory, Chinese Academy of Sciences, 10 Yuanhua Road, Qixia District,
Nanjing 210023, China}
\email{libin@pmo.ac.cn}
\author{Lin Nie}
\affiliation{Shanghai Astronomical Observatory, Chinese Academy of Sciences, Shanghai 200030, People's Republic of China}
\email{linn@shao.ac.cn}
\author{Zhaojun Yan}
\affiliation{Shanghai Astronomical Observatory, Chinese Academy of Sciences, Shanghai 200030, People's Republic of China}
\email{zhaojunyan@shao.ac.cn}
\author{Xuechun Chen}
\affiliation{Beijing City University, No. 26 Muyan Road (Yangzhen Section), Yangzhen, Shunyi District, Beijing 101309, People's Republic of China}
\email{xuechun93@foxmail.com}
\author{Yachong Diao}
\affiliation{Beijing City University, No. 26 Muyan Road (Yangzhen Section), Yangzhen, Shunyi District, Beijing 101309, People's Republic of China}
\email{diaoyachong@gmail.com}
\author{Huanyuan Shan}
\affiliation{Shanghai Astronomical Observatory, Chinese Academy of Sciences, Shanghai 200030, People's Republic of China}
\email{hyshan@shao.ac.cn}

\begin{abstract}
We present Finite-iteration Deconvolution--Reconvolution Denoising (FiDeR), a method for reducing noise in astronomical images while retaining approximately the input point-spread function (PSF). We first apply a finite number of Richardson--Lucy (RL) iterations to the nonnegative source-plus-background image. We then measure the response of a calibration point source processed with the same settings and derive a reconvolution kernel matched to the observational PSF. The calibration includes the background-to-source flux ratio. We compare FiDeR with six denoisers using simulated point-source and mixed stellar--galaxy fields, and observations from the James Webb Space Telescope Near-Infrared Camera (NIRCam) and Mid-Infrared Instrument (MIRI). In the stellar tests, bright-source widths remain close to the input widths; faint sources show small residual offsets. For the mixed field, FiDeR reaches peak signal-to-noise ratios of 46.701 and 44.850 dB at backgrounds of 10 and 20 counts per pixel, with wing errors of 0.657 and 0.858 counts. It also gives small aperture-flux and shape errors, although block-matching and three-dimensional filtering (BM3D) gives higher structural similarity. The NIRCam data show lower core and combined-region errors, while the three-exposure MIRI test gives benefits mainly in the stellar aperture and photometry. These results support substantial noise reduction with little change in the measured PSF within the tested regimes. The denoised images can provide inputs for further reconstruction and measurement; their uncertainties must account for residual noise correlations.
\end{abstract}

\section{Introduction}\label{sec:intro}
Astronomical image processing must reduce noise without unduly changing the source profile. A smoother background is useful for detecting faint structure, but changes in the point-spread function (PSF) can affect aperture photometry, morphology, and subsequent image reconstruction. We therefore seek a denoised image at the observational resolution, rather than the sharpest possible reconstruction.

Local mean and median filters provide simple noise-reduction methods. A mean filter introduces an additional convolution, while a median can alter source peaks and enclosed flux. Adaptive filters and wavelet methods adjust their response to local structure or scale \citep{lee1980,donoho1995}. Non-local means (NLM) and block-matching and three-dimensional filtering (BM3D) use similarities between image patches to suppress noise \citep{buades2005,buades2011,dabov2007}. These methods can retain source structure better than fixed smoothing, but do not explicitly match the output to a prescribed astronomical PSF.

Richardson--Lucy (RL) reconstruction uses a known PSF and a Poisson likelihood to recover a nonnegative image \citep{richardson1972,lucy1974}. Stopping the iteration before convergence provides regularization \citep{prato2012}, but also leaves a finite-iteration point-source response. Reconvolving the reconstruction can return it to a chosen resolution. Restoration followed by reblurring, controlled-resolution deconvolution, and common-resolution kernels are established approaches \citep{kawata1980,magain1998,aniano2011}. The response of a finite nonlinear iteration, however, need not be a delta function or be independent of the background and source flux.

We introduce Finite-iteration Deconvolution--Reconvolution Denoising (\FiDeR). We calibrate the point-source response of the finite RL reconstruction at the relevant background-to-source ratio and use that response to derive the reconvolution kernel. The target is normally the input PSF. We reconstruct source and background together, then subtract the background after reconvolution. The method extends our use of PSF-based iterative processing in upsampling and PSF deconvolution coaddition (UPDC; \citealt{wang2022updc}), UPdec-Webb \citep{wang2025updec}, and filtered image deconvolution and noise backfilling (FiDNoBa; \citealt{zhang2026finoba}) to a different task: denoising at the observational resolution.

We test \FiDeR\ against six denoisers in point-source and mixed stellar--galaxy simulations, then apply the same approach to James Webb Space Telescope (JWST) observations. We measure image errors, stellar widths, aperture fluxes, and shapes separately. Section~\ref{sec:estimator} describes the method, Sections~\ref{sec:simulations} and \ref{sec:observations} present the tests, and Section~\ref{sec:conclusions} summarizes the results and applications. The appendices give the diagnostics, conservation proof, and noise-treatment details.

\section{The FiDeR Method}\label{sec:estimator}
\subsection{Finite-iteration reconstruction}
Let $f_i\geq0$ be the source intensity before convolution and $H$ the convolution operator for a nonnegative, unit-sum PSF $h$. For a locally constant Poisson background $b\geq0$,
\begin{equation}
y_i\sim\operatorname{Poisson}(\mu_i),\qquad \mu_i=(Hf)_i+b.\label{eq:poisson}
\end{equation}
Our target is the source image $s=Hf$ at the observational resolution. With $H\mathbf{1}=\mathbf{1}$, the total latent intensity $x=f+b\mathbf{1}$ satisfies $\mu=Hx$. We reconstruct this nonnegative source-plus-background intensity. For a background-subtracted input $E=y-b\mathbf{1}$, we first restore $b$; subtracting the mean background does not remove its noise. A spatially varying background requires a separate model and cannot in general be moved through $H$ in this way.

The PSF must include the pixel-sampling response of the working grid. We use the Poisson negative log likelihood
\begin{equation}
\mathcal{L}(x;y)=\sum_i[(Hx)_i-y_i\log(Hx)_i],\qquad x_i\geq0,
\end{equation}
and the RL update
\begin{equation}
x^{(j+1)}=x^{(j)}\odot H^\mathsf{T}\left[\frac{y}{Hx^{(j)}}\right],
\qquad x^{(0)}=y.\label{eq:rl}
\end{equation}
Products and quotients are elementwise, and $H^\mathsf{T}$ is the adjoint. Equation~\ref{eq:rl} assumes complete support and $H^\mathsf{T}\mathbf{1}=\mathbf{1}$; otherwise the update includes the sensitivity normalization $H^\mathsf{T}\mathbf{1}$. Small positive floors protect initialization and division. We denote the result after $k$ iterations by $\RL_k(y;h)$. The tests use iteration budgets of 200 or 1000. We apply identical iteration counts, initialization, PSF discretization, boundary treatment, and numerical floors to the image and its response calibration.

\subsection{Point-source response and matching kernel}
We process a noiseless calibration point source of flux $F_\mathrm{c}$ on the same background, $F_\mathrm{c}h+b\mathbf{1}$, and define its normalized background-subtracted response as
\begin{equation}
K_{e,k}^{(\beta)}=
\frac{\RL_k(F_\mathrm{c}h+b\mathbf{1};h)-b\mathbf{1}}{F_\mathrm{c}}
=\RL_k(h+\beta\mathbf{1};h)-\beta\mathbf{1},\qquad
\beta=\frac{b}{F_\mathrm{c}}.\label{eq:effective}
\end{equation}
The second equality follows from RL homogeneity before fixed numerical floors are imposed. It shows why the background-to-source ratio enters the calibration. The response sums to unity under the conservation assumptions in Appendix~\ref{app:conservation}, but background subtraction can leave negative values.

We use $K_{e,k}^{(\beta)}$ to describe this particular finite reconstruction, rather than a global impulse response of the nonlinear estimator. It is also distinct from an effective PSF that incorporates detector sampling \citep{anderson2000epsf}. The calibration flux is known in simulations and estimated from a suitable source in observations. Different source fluxes can have different responses, so the adopted $F_\mathrm{c}$ is part of the estimator specification.

We seek a kernel $a_k$ such that $a_k*K_{e,k}^{(\beta)}\simeq h_\star$, where the target $h_\star$ is normally the input PSF $h$. A regularized Fourier solution is
\begin{equation}
\widehat a_k(\boldsymbol\nu)=
\frac{\widehat K_{e,k}^{(\beta)*}(\boldsymbol\nu)\widehat h_\star(\boldsymbol\nu)}
{|\widehat K_{e,k}^{(\beta)}(\boldsymbol\nu)|^2+\eta},
\qquad \widehat a_k(\mathbf{0})=1.\label{eq:matching}
\end{equation}
Hats denote discrete Fourier transforms and the asterisk denotes complex conjugation. We use $\eta=3\times10^{-14}$ in the normalized discrete convention and explicitly set the zero-frequency coefficient to unity. The final source image is
\begin{equation}
I_k=a_k*\RL_k(y;h)-b\mathbf{1}
=a_k*\RL_k(E+b\mathbf{1};h)-b\mathbf{1}.\label{eq:estimator}
\end{equation}
The matching kernel can have negative lobes. We retain negative output pixels to avoid imposing a positive background bias or changing the measured flux.

For the calibration source, the nonzero-frequency matching error is
\begin{equation}
\widehat a_k\widehat K_{e,k}^{(\beta)}-\widehat h_\star
=-\widehat h_\star\frac{\eta}{|\widehat K_{e,k}^{(\beta)}|^2+\eta}.\label{eq:matcherror}
\end{equation}
The unregularized solution matches the target wherever the response spectrum is nonzero. Near spectral zeros, regularization is necessary. The calibration residual should be checked together with the kernel's maximum spectral gain; a small residual alone does not establish stability for noisy data.

\subsection{Flux conservation and implementation}
The normalized RL update conserves total counts on a complete working domain. The unit zero-frequency coefficient preserves them during reconvolution, giving
\begin{equation}
\sum_i I_{k,i}=\sum_i y_i-Pb=\sum_i E_i,\label{eq:flux}
\end{equation}
where $P$ is the number of pixels. This statement applies to the full domain under the assumptions in Appendix~\ref{app:conservation}. Finite-aperture flux and morphology still require measurement because the processing moves light across aperture boundaries.

In practice, we normalize and center the sampled PSF, estimate $b$ and $F_\mathrm{c}$, calibrate the response and matching kernel, and then reconstruct and reconvolve the image. Fast Fourier transform (FFT) convolution gives a cost of approximately $\mathcal{O}(kP\log P)$. A calibration can be reused while the PSF, $\beta$, and iteration settings remain unchanged. Image and calibration convolutions use the same padding and boundary convention. Appendix~\ref{app:noise} describes calibrated intensity units, non-Poisson backgrounds, and residual covariance.

\section{Simulation Experiments}\label{sec:simulations}
\subsection{Experimental design}
We test a crowded point-source field and a mixed stellar--galaxy field. We convolve each source model with a normalized Gaussian PSF, add a uniform background of $b=10$ or 20 counts per pixel, and Poisson sample the complete image. Both source and background shot noise are therefore present. All methods receive the same realization. We subtract the known mean background before measuring the source image; FiDeR restores it during the RL stage.

We compare FiDeR with six denoisers: $3\times3$ mean and median filters, a local spatial Kalman update, Haar shrinkage, non-local means (NLM), and BM3D. The Kalman update uses the mean and variance $v$ of the eight neighboring pixels and a gain $v/(v+\sigma_n^2)$. Haar uses two levels, eight shifted decompositions, and smooth attenuation of detail coefficients. NLM uses $h=1.1\widehat\sigma$, $17\times17$ patches, and a 25-pixel search distance on the original intensity scale. BM3D uses the official package with a common normalization of the image and noise parameter. These comparisons use the implementations and settings specified in Appendix~\ref{app:implementation}.

We measure image root mean square error (RMSE), peak signal-to-noise ratio (PSNR), structural similarity (SSIM; \citealt{wang2004}), wing RMSE, aperture-flux errors, shape errors, and fitted stellar widths. The reference is always the noiseless \emph{PSF-convolved} scene. Apertures and centers are fixed across realizations and methods. The aperture reference includes neighboring light, so these flux errors describe blended apertures. Shape moments use the known source positions; position-angle errors exclude nearly circular reference profiles with distortion $\chi<0.05$. Stellar widths use a circular-Gaussian-plus-constant fit within $r\leq8$ pixels at the true center. We give the metric definitions and sampling conventions in Appendix~\ref{app:diagnostics}.

\subsection{Mock-I: 36 point sources}\label{sec:mock1}
\begin{figure*}[htbp]
\centering\includegraphics[width=\textwidth]{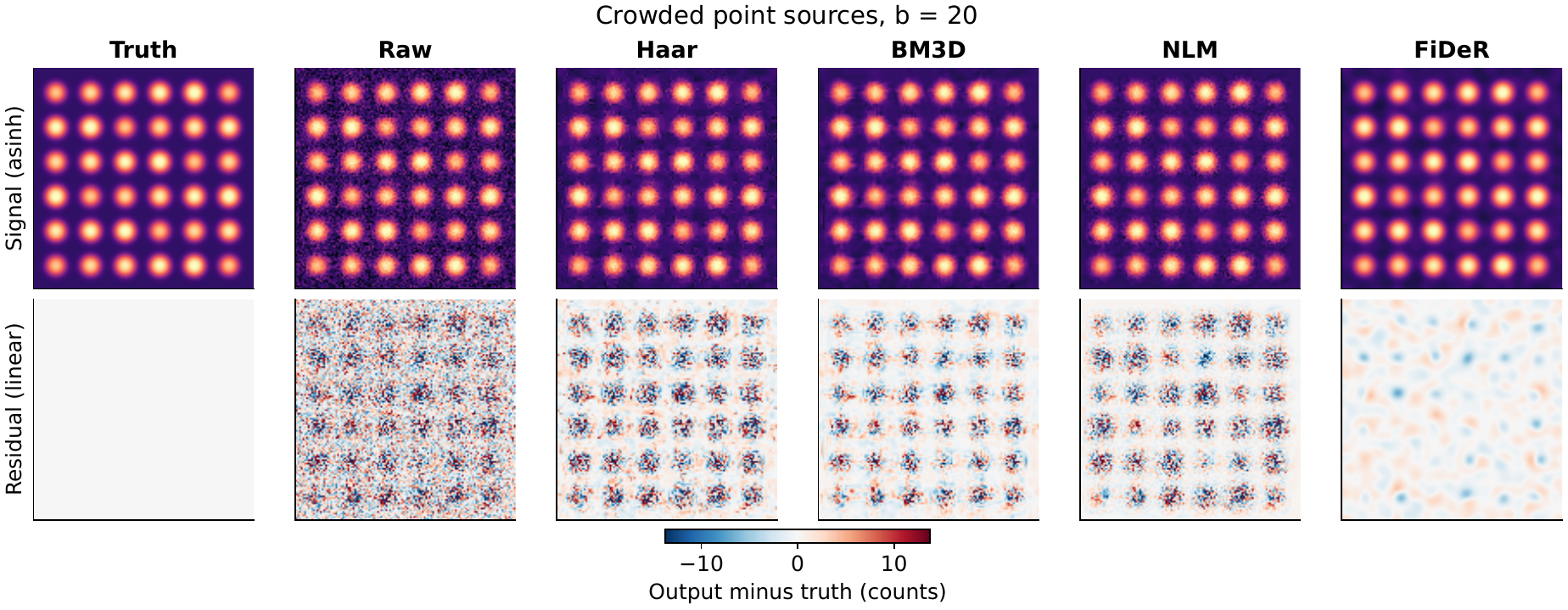}
\caption{Mock-I: 36 stars at $b=20$ counts per pixel. Upper row: the noiseless PSF-convolved scene, raw image, and selected denoisers on a common asinh scale. Lower row: output minus the same noiseless scene on a common linear count scale. FiDeR reduces pixel fluctuations but leaves low-amplitude spatial structure. Figure~\ref{fig:widths} tests the corresponding width behavior.}\label{fig:crowded}
\end{figure*}

The crowded field contains a $6\times6$ grid on a $128\times128$ image, with 20-pixel separations and source fluxes of 6000--11,280 counts. The Gaussian PSF has a nominal full width at half maximum (FWHM) of 7 pixels and a normalized $31\times31$ support. At about 2.9 FWHM separation, neighboring wings overlap. Figure~\ref{fig:crowded} shows the $b=20$ case. FiDeR suppresses the visible pixel fluctuations; its residuals are mainly coherent, low-amplitude variations rather than high-frequency noise.

We then isolate the width response with equal-flux fields. Each field contains 36 stars with either 3000 or 30,000 counts per star. FWHMs of 4, 6, and 8 pixels, combined with the two backgrounds, give 12 cases with 100 independent Poisson realizations each. We enlarge the image to $160\times160$ pixels; centers run from 29 to 129 pixels at 20-pixel spacing. This test shares the geometry of the crowded field but uses different fluxes, PSFs, and edge margins.

\begin{figure*}[htbp]
\centering\includegraphics[width=\textwidth]{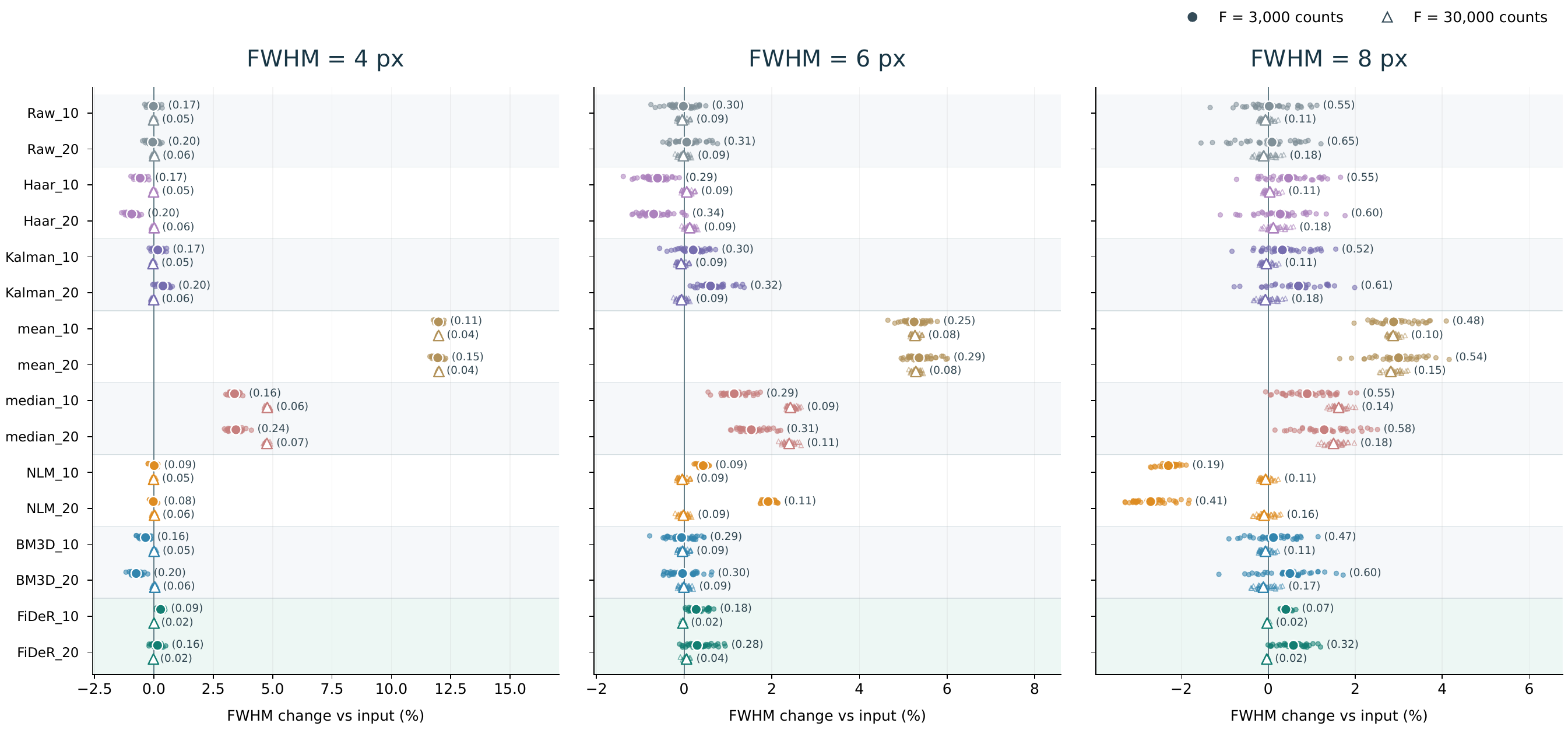}
\caption{Equal-flux Mock-I width test. Columns show nominal FWHMs of 4, 6, and 8 pixels. Suffixes 10 and 20 denote background counts per pixel; filled circles and open triangles denote 3000 and 30,000 counts per star. Small symbols show individual stellar widths measured from the Monte Carlo mean images relative to identically fitted noiseless comparators; they represent stars, not independent realizations. Parentheses give the sample scatter across 36 stars in percentage points. Larger symbols summarize each stellar distribution.}\label{fig:widths}
\end{figure*}

The mean filter broadens every tested PSF, with the largest relative change at 4 pixels. A $3\times3$ box adds $2/3$ pixel$^2$ to each coordinate variance in an ideal moment calculation. Median filtering also shifts the fitted widths, while the low-flux NLM results vary with width and background and include negative offsets for some broader PSFs. Haar and Kalman keep widths closer to the reference, although the crowded-field residual maps retain source-scale fluctuations.

FiDeR's bright-source width changes remain close to zero, with scatters of 0.02--0.04 percentage points. For the low-flux fields the scatters are 0.07--0.32 percentage points, and some distributions have positive offsets of several tenths of a percent. FiDeR and BM3D are among the two best-performing methods in this width test. These results support approximate width preservation, particularly for bright sources. They also show a residual dependence on source strength. Since several methods preserve bright-source FWHM, we assess width together with residual noise and photometry; FWHM alone does not characterize the PSF wings or asymmetry. The equal-flux scatter measures star-to-star variation within a field. Dependence on flux follows from comparing the two flux groups, rather than from that scatter itself.

\subsection{Mock-II: stars and galaxies}\label{sec:mock2}
The mixed scene places eight stars and eight galaxies on an interleaved $4\times4$ grid in a $128\times128$ image. We use the same nominal 7-pixel Gaussian PSF as in the crowded-field visual test. Stellar fluxes span 7000--10,000 counts. The galaxies are finite-image-normalized S\'ersic models with total flux 10,000 counts, indices $n=1,2,4$, major-to-minor axis ratios 2, 4, and 7, and multiple orientations. This scene tests compact cores, extended wings, and anisotropic profiles together.

\begin{figure*}[htbp]
\centering\includegraphics[width=\textwidth]{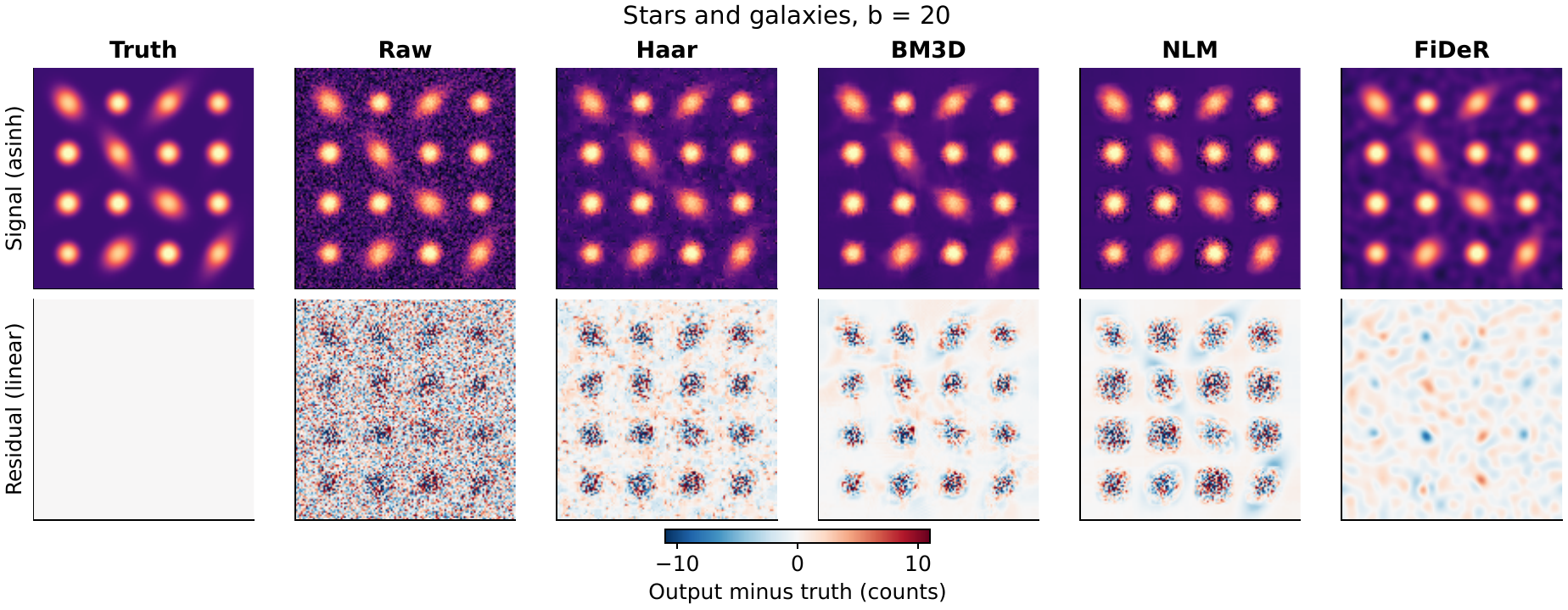}
\caption{Mock-II at $b=20$ counts per pixel. Signal and residual rows use the same conventions as Figure~\ref{fig:crowded}. Stars and elliptical galaxies are interleaved. FiDeR suppresses pixel-scale fluctuations while retaining the main source geometries. Table~\ref{tab:mock2} gives the image, wing, flux, ellipticity, and position-angle diagnostics for both backgrounds.}\label{fig:mixed}
\end{figure*}

FiDeR reduces fluctuations around both compact and extended sources in Figure~\ref{fig:mixed}. Its residuals remain spatially structured, with low-frequency variations across the field. BM3D and NLM also leave coherent residuals, particularly near source outskirts. FiDeR does not apply separate rules to high- and low-S/N regions.

\begin{table*}[htbp]
\centering
\caption{Mixed stellar--galaxy simulation at backgrounds $b=10$ and $b=20$ counts per pixel. Entries are means of $N_{\rm MC}=200$ realizations per level. Arrows indicate the preferred direction; bold values denote the best result in each column.}
\label{tab:mock2}
\setlength{\tabcolsep}{3.2pt}
\renewcommand{\arraystretch}{1.1}
\begin{tabular}{lcccccc|cccccc}
\toprule
& \multicolumn{6}{c|}{\textbf{$b=10$}}
& \multicolumn{6}{c}{\textbf{$b=20$}} \\
\cmidrule(lr){2-7}
\cmidrule(lr){8-13}

Method
& PSNR & SSIM & Wing & Phot. & Ellip. & PA
& PSNR & SSIM & Wing & Phot. & Ellip. & PA \\
& (dB)$\uparrow$
& $\uparrow$
& RMSE$\downarrow$
& MAE(\%)$\downarrow$
& Err.$\downarrow$
& Err.($^\circ$)$\downarrow$
& (dB)$\uparrow$
& $\uparrow$
& RMSE$\downarrow$
& MAE(\%)$\downarrow$
& Err.$\downarrow$
& Err.($^\circ$)$\downarrow$ \\
\midrule

Raw
& 32.306 & 0.7761 & 3.616 & 0.97 & 0.0117 & 1.316
& 30.468 & 0.6626 & 4.805 & 1.08 & 0.0130 & 1.706 \\

Haar
& 34.844 & 0.9276 & 2.001 & 1.07 & 0.0114 & 1.253
& 34.250 & 0.8928 & 2.258 & 1.22 & 0.0124 & 1.585 \\

Kalman
& 34.427 & 0.8931 & 2.287 & 0.97 & 0.0114 & 1.259
& 33.480 & 0.8328 & 2.774 & 1.08 & 0.0124 & 1.602 \\

Mean
& 39.702 & 0.9427 & 1.284 & 1.07 & 0.0146 & 1.227
& 38.491 & 0.9022 & 1.659 & 1.20 & 0.0144 & 1.533 \\

Median
& 37.000 & 0.9164 & 1.605 & 3.86 & 0.0120 & 1.313
& 35.963 & 0.8642 & 2.061 & 3.65 & 0.0131 & 1.652 \\

NLM
& 34.480 & 0.9297 & 2.423 & 0.99 & 0.0112 & 1.227
& 34.578 & 0.9261 & 2.352 & 1.23 & 0.0119 & 1.497 \\

BM3D
& 36.190 & \textbf{0.9671} & 1.385 & 1.04 & 0.0111 & 1.232
& 36.549 & \textbf{0.9592} & 1.403 & 1.18 & 0.0125 & 1.580 \\

FiDeR
& \textbf{46.701} & 0.9670 & \textbf{0.657} & \textbf{0.96} & \textbf{0.0107} & \textbf{1.186}
& \textbf{44.850} & 0.9408 & \textbf{0.858} & \textbf{1.06} & \textbf{0.0115} & \textbf{1.463} \\

\bottomrule
\end{tabular}
\end{table*}

Table~\ref{tab:mock2} averages 200 realizations at each background. FiDeR gives PSNRs of 46.701 and 44.850 dB at $b=10$ and 20, exceeding BM3D by 10.511 and 8.301 dB. Wing RMSE falls to 0.657 and 0.858 counts, compared with BM3D's 1.385 and 1.403 counts. We measure these wings in the union of source-centered annuli $7\leq r\leq14$ pixels. The improvement therefore extends beyond the brightest pixels.

FiDeR's aperture-flux MAEs are 0.96\% and 1.06\%, compared with 1.04\% and 1.18\% for BM3D. Aperture radii follow the noiseless annular source S/N threshold, with a maximum of 14.5 pixels. FiDeR also has the smallest reported ellipticity and position-angle errors, although the differences among the leading methods are small and the table provides ensemble means without paired uncertainties. Median filtering produces larger flux errors, including 3.65\% at $b=20$.

BM3D performs better in SSIM. At $b=10$, its 0.9671 and FiDeR's 0.9670 are nearly equal at the displayed precision. At $b=20$, the scores are 0.9592 and 0.9408. Mean filtering also illustrates the difference between diagnostics: its PSNR of 38.491 dB exceeds BM3D's 36.549 dB at $b=20$, but its SSIM is lower, 0.9022, and it broadens the stellar response. The tests therefore favor FiDeR for pixel and wing error reduction with small width and aperture changes, while identifying regimes and metrics where other denoisers perform better.

\FloatBarrier
\section{JWST Observations}\label{sec:observations}
\subsection{References and measurements}
We test nine Near-Infrared Camera (NIRCam) exposures of the SMACS J0723 field, using F200W PSF information, and three Mid-Infrared Instrument (MIRI) F1800W exposures with a common empirical PSF for the selected cutouts. The two data sets sample different PSF widths and detector-sampling regimes. Unlike the simulations, their references retain instrumental structure, calibration residuals, and noise.

We register the exposures on a common grid, use compatible intensity units and background subtraction, and construct Drizzle-based coadds \citep{fruchter2002}. For resampled exposures $d_j$ and weights $w_j$, the full and leave-one-out references are
\begin{equation}
T=\frac{\sum_j w_jd_j}{\sum_jw_j},\qquad
T_{-i}=\frac{\sum_{j\ne i}w_jd_j}{\sum_{j\ne i}w_j}.\label{eq:coadds}
\end{equation}
These are mean intensities on a common flux scale; equal weights give arithmetic means. We calculate the primary statistics for individual processed exposures, then average the measurements. Averaged output images provide a separate visual comparison.

For each NIRCam exposure, we use the other eight raw exposures as the reference. For MIRI, we use the three-exposure coadd, which includes the tested exposure and therefore correlates its noise with the residual. Neither reference is noiseless, and its resampling response can differ from that of a native detector image. PSF matching uses the working sampling.

We process background-subtracted inputs as in Section~\ref{sec:estimator}. For calibrated cutouts without a physical count scale, we convert the intensities to effective counts and restore a variance-matched background pedestal as described in Appendix~\ref{app:noise}. This approximates the noise amplitude without assuming independent Poisson statistics after resampling.

\subsection{Observation-I: NIRCam}
\begin{figure*}[htbp!]
\centering
\includegraphics[width=0.97\textwidth]{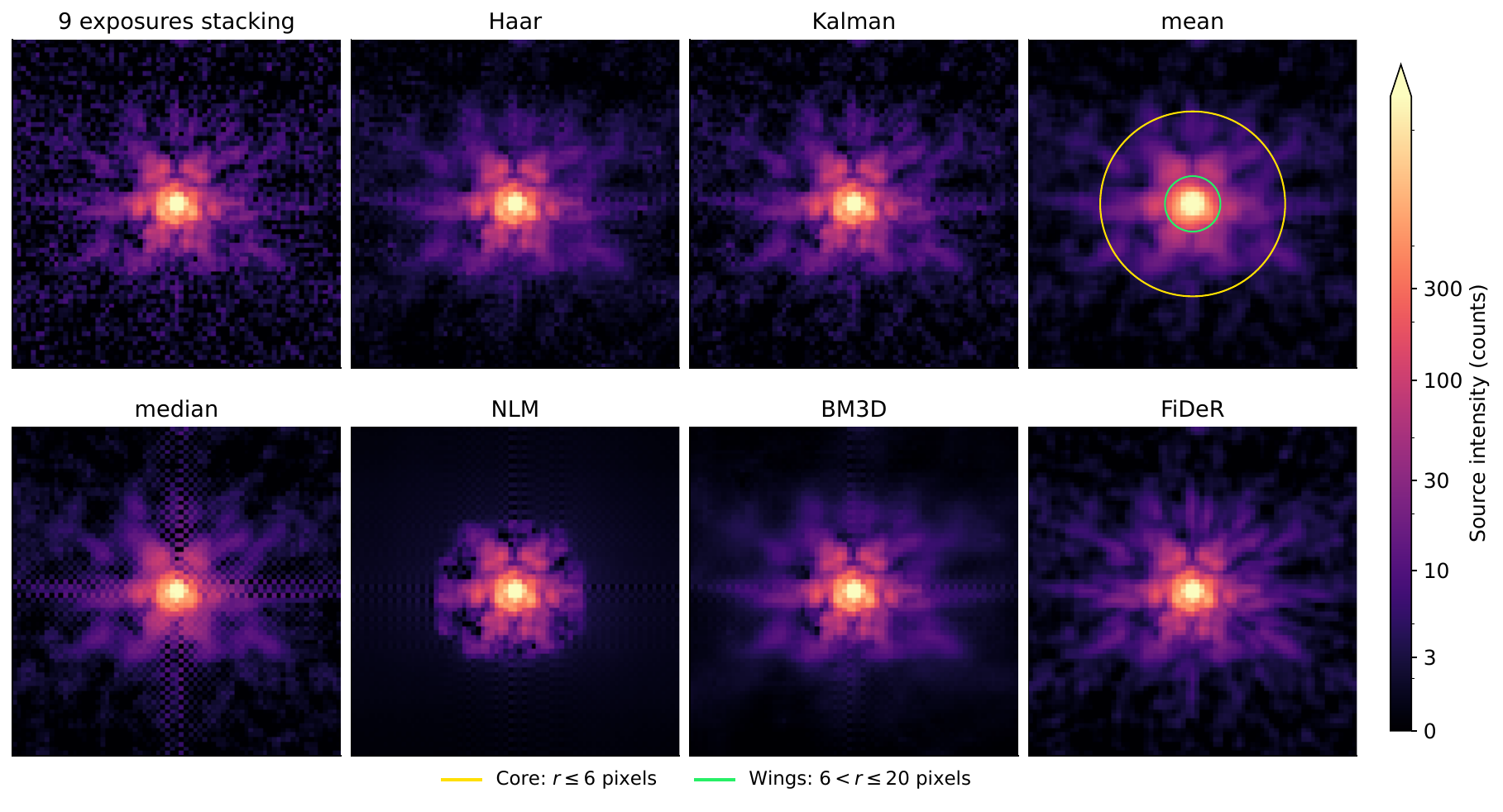}\par\vspace{2mm}
\includegraphics[width=0.97\textwidth]{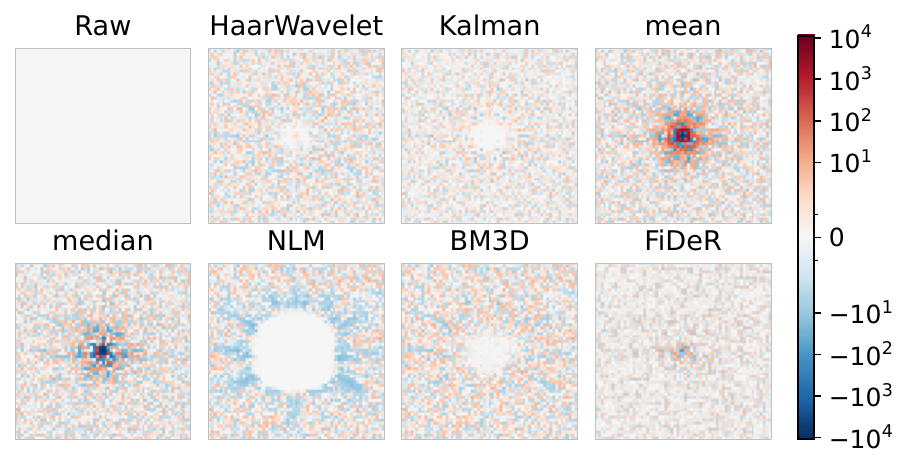}
\caption{NIRCam comparison. Top: nine-exposure coadd and mean denoised images, with a stretch emphasizing faint structure and circular boundaries at $r=6$ and 20 pixels. Bottom: corresponding residual comparison on a common symmetric-log scale in counts. The bottom ``Raw'' panel is the reference coadd subtracted from itself and is zero by construction. HaarWavelet and Haar denote the same method. Each block retains its own crop and display scale. The average-output residuals differ from the single-exposure leave-one-out statistic in Table~\ref{tab:nircam}.}\label{fig:nircam}
\end{figure*}

Figure~\ref{fig:nircam} shows the nine-exposure reference, averaged denoised images, and residual maps. The intensity stretch emphasizes faint structure. NLM smooths some weak stellar-wing features, whereas FiDeR leaves little coherent source structure in the displayed residuals.

\begin{table}[htbp]
\centering\small
\caption{Observation-I: mean single-exposure errors relative to a leave-one-out NIRCam reference.}
\label{tab:nircam}
\setlength{\tabcolsep}{5pt}
\renewcommand{\arraystretch}{1.12}

\begin{tabular}{lrrrr}
\toprule
Method & Core RMSE & Wing RMSE & Core+wing RMSE & Core+wing SSIM \\
 & $r\leq6$ & $6<r\leq20$ & $r\leq20$ & $r\leq20$ \\
 & (counts)$\downarrow$ & (counts)$\downarrow$ & (counts)$\downarrow$ & $\uparrow$ \\
\midrule

Raw
& 35.366
& 8.567
& 13.408
& 0.9997065 \\

Haar
& 35.587
& 6.165
& 12.195
& 0.9997597 \\

Kalman
& 35.349
& 6.431
& 12.260
& 0.9997586 \\

Mean
& 1313.263
& 9.819
& 393.864
& 0.9658530 \\

Median
& 1514.916
& 10.203
& 454.318
& 0.9479879 \\

NLM
& 35.366
& 7.449
& 12.781
& 0.9997104 \\

BM3D
& 35.304
& \textbf{5.599}
& 11.867
& 0.9997806 \\

FiDeR
& \textbf{16.598}
& 5.735
& \textbf{7.411}
& \textbf{0.9998485} \\

\bottomrule
\end{tabular}

\par\smallskip
\parbox{0.97\linewidth}{\footnotesize
Entries are arithmetic means of nine comparisons, each against the mean
reference formed from the other eight exposures. The core is $r\leq6$ pixels
and the wings are $6<r\leq20$ pixels. Combined SSIM uses 573 centers whose
complete $11\times11$ Gaussian window lies within $r\leq20$, with window
standard deviation 1.5 pixels and population covariance. The dynamic range
is measured inside the corresponding reference aperture.}
\end{table}

We define the core by $r\leq6$ pixels and the wings by $6<r\leq20$ pixels. Table~\ref{tab:nircam} averages the nine leave-one-out comparisons. Combined RMSE uses all pixels inside $r\leq20$ directly; SSIM uses the 573 centers whose complete windows lie inside this aperture.

FiDeR reduces core RMSE to 16.598 counts, from 35.366 for Raw and 35.304 for BM3D. Its combined RMSE is 7.411 counts, compared with 13.408 and 11.867. Relative to BM3D, these reductions are approximately 53\% and 38\%. Mean and median filtering instead give core RMSEs of 1313.263 and 1514.916 counts, consistent with substantial changes to the bright central profile.

BM3D has the lowest wing RMSE, 5.599 counts, slightly below FiDeR's 5.735; both improve on Raw's 8.567. Haar and Kalman give 6.165 and 6.431 counts. NLM gives 7.449 counts in the wings and 12.781 counts overall, consistent with its loss of weak structure in this cutout. FiDeR has the highest combined SSIM, 0.9998485, compared with 0.9997806 for BM3D. Both scores are near unity, and the table does not provide uncertainties on their difference. The main NIRCam advantage is the reduction in core and combined residuals.

\subsection{Observation-II: MIRI}
\begin{figure*}[htbp!]
\centering\includegraphics[width=0.98\textwidth,height=0.78\textheight,keepaspectratio]{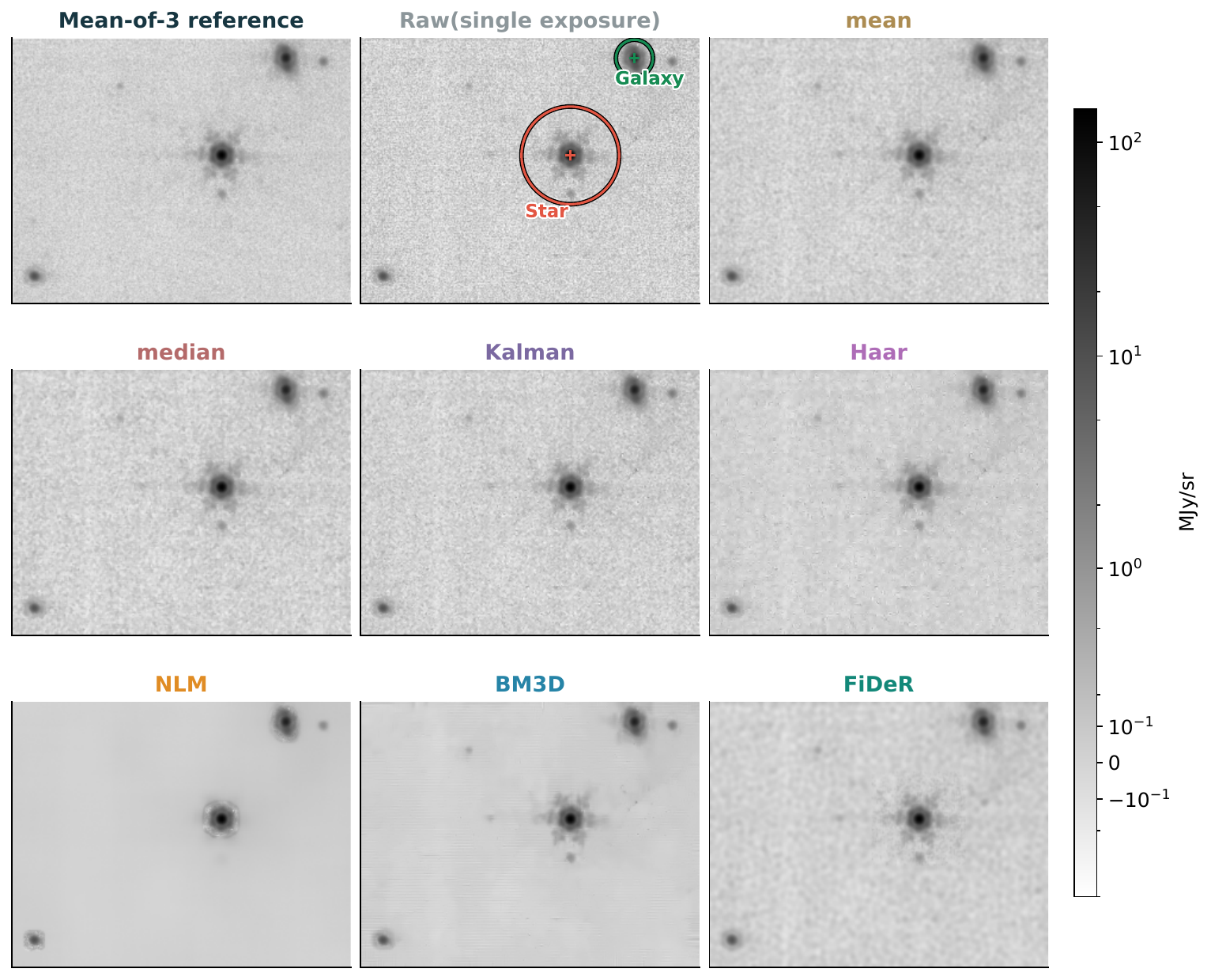}
\caption{MIRI comparison. The first panel is the three-exposure reference; the second is one raw exposure, followed by its denoised counterparts. All panels use a common surface-brightness scale in MJy sr$^{-1}$. Tables~\ref{tab:miri_metrics} and \ref{tab:miri_photometry} report means and sample standard deviations of three individual-exposure measurements, rather than measurements of this displayed exposure alone.}\label{fig:miri}
\end{figure*}

Figure~\ref{fig:miri} shows the three-exposure reference and one raw exposure with its processed counterparts. These outputs are single exposures, unlike the NIRCam average images. The stellar profile is broader in detector pixels than in the NIRCam example.

We measure the common image footprint and fixed apertures around a star at $(\alpha,\delta)=(215.1734178^\circ,53.0414361^\circ)$ and a galaxy at $(215.1793158^\circ,53.0406626^\circ)$. Their aperture radii are 50 and 19 pixels. The stellar aperture contains neighboring light. The full footprint, stellar aperture, and galaxy aperture contain 93,420, 7858, and 1132 pixels; their SSIM supports contain 87,360, 5958, and 483 complete-window centers. The galaxy lies near the cutout boundary and remains sensitive to the assumed exterior image even though its aperture is fully covered.

\begin{table*}[htbp!]
\centering
\caption{Observation-II: image-quality metrics for the three MIRI exposures.}
\label{tab:miri_metrics}

\begingroup
\footnotesize
\setlength{\tabcolsep}{5pt}
\renewcommand{\arraystretch}{1.02}

\begin{center}
\textbf{Full image (common coverage)}\par\smallskip

\begin{tabular}{lrrr}
\toprule
Method
& RMSE [MJy/sr]$\downarrow$
& PSNR [dB]$\uparrow$
& SSIM$\uparrow$ \\
\midrule

Raw
& $0.146 \pm 0.031$
& $59.991 \pm 1.751$
& $0.998356 \pm 0.000500$ \\

Mean
& $0.194 \pm 0.012$
& $57.443 \pm 0.567$
& $0.998909 \pm 0.000069$ \\

Median
& $0.171 \pm 0.003$
& $58.487 \pm 0.145$
& $0.998604 \pm 0.000129$ \\

Kalman
& $0.104 \pm 0.007$
& $62.811 \pm 0.574$
& $0.998995 \pm 0.000093$ \\

Haar
& $\mathbf{0.100 \pm 0.005}$
& $\mathbf{63.132 \pm 0.473}$
& $0.999139 \pm 0.000081$ \\

NLM
& $0.113 \pm 0.002$
& $62.080 \pm 0.164$
& $0.998693 \pm 0.000040$ \\

BM3D
& $0.104 \pm 0.001$
& $62.801 \pm 0.121$
& $\mathbf{0.999153 \pm 0.000056}$ \\

FiDeR
& $0.112 \pm 0.005$
& $62.215 \pm 0.358$
& $0.999037 \pm 0.000046$ \\

\bottomrule
\end{tabular}
\end{center}

\endgroup

\par\medskip

\begingroup
\footnotesize
\setlength{\tabcolsep}{4.0pt}
\renewcommand{\arraystretch}{1.08}

\begin{center}

\begin{tabular}{lrrr|rrr}
\toprule
& \multicolumn{3}{c|}{\textbf{Star: $r\leq50$ pixels}}
& \multicolumn{3}{c}{\textbf{Galaxy: $r\leq19$ pixels}} \\
\cmidrule(lr){2-4}
\cmidrule(lr){5-7}

Method
& RMSE [MJy/sr]$\downarrow$
& PSNR [dB]$\uparrow$
& SSIM$\uparrow$
& RMSE [MJy/sr]$\downarrow$
& PSNR [dB]$\uparrow$
& SSIM$\uparrow$ \\

\midrule

Raw
& $0.165 \pm 0.037$
& $58.939 \pm 1.864$
& $0.998276 \pm 0.000415$
& $0.172 \pm 0.036$
& $49.067 \pm 1.765$
& $0.996338 \pm 0.001783$ \\

Mean
& $0.534 \pm 0.063$
& $48.650 \pm 1.069$
& $0.998332 \pm 0.000122$
& $0.462 \pm 0.067$
& $40.431 \pm 1.266$
& $0.993094 \pm 0.001654$ \\

Median
& $0.422 \pm 0.027$
& $50.662 \pm 0.556$
& $0.998435 \pm 0.000137$
& $0.366 \pm 0.042$
& $42.451 \pm 1.018$
& $0.995672 \pm 0.000601$ \\

Kalman
& $0.131 \pm 0.009$
& $60.804 \pm 0.568$
& $0.998882 \pm 0.000128$
& $0.151 \pm 0.016$
& $50.130 \pm 0.940$
& $0.997208 \pm 0.000664$ \\

Haar
& $0.129 \pm 0.011$
& $60.943 \pm 0.696$
& $0.998946 \pm 0.000105$
& $0.152 \pm 0.019$
& $50.061 \pm 1.090$
& $0.997103 \pm 0.000680$ \\

NLM
& $0.172 \pm 0.013$
& $58.481 \pm 0.631$
& $0.994242 \pm 0.000360$
& $0.157 \pm 0.025$
& $49.818 \pm 1.367$
& $0.996553 \pm 0.001566$ \\

BM3D
& $0.129 \pm 0.011$
& $60.952 \pm 0.690$
& $0.998963 \pm 0.000102$
& $\mathbf{0.138 \pm 0.013}$
& $\mathbf{50.873 \pm 0.834}$
& $\mathbf{0.997948 \pm 0.000105}$ \\

FiDeR
& $\mathbf{0.123 \pm 0.016}$
& $\mathbf{61.405 \pm 1.077}$
& $\mathbf{0.998994 \pm 0.000040}$
& $0.155 \pm 0.012$
& $49.860 \pm 0.685$
& $0.997489 \pm 0.000651$ \\

\bottomrule
\end{tabular}

\end{center}
\endgroup

\par\smallskip

\parbox{0.97\linewidth}{\footnotesize
Entries are means and sample standard deviations across three exposures,
each compared with the three-exposure reference. Full-image measurements
use common coverage; stellar and galaxy radii are 50 and 19 pixels,
respectively. SSIM uses complete windows inside the corresponding region.
Standard deviations are not standard errors, and the reference includes
the tested exposure. Boldface indicates the best value among the denoising
methods; the Raw row is treated as the unfiltered reference and is not
included in the ranking. Rankings are determined using the original
unrounded values.}

\end{table*}

FiDeR reduces full-image RMSE from Raw's 0.1461 to 0.1116 MJy sr$^{-1}$ and raises PSNR from 59.991 to 62.215 dB. Haar has the lowest full-image RMSE, 0.1004, while BM3D gives 0.1043 and the highest SSIM, 0.999153; FiDeR gives 0.999037. Thus FiDeR improves the raw image but does not lead the full-field comparison.

Within the stellar aperture, FiDeR gives the lowest RMSE, $0.1230\pm0.0157$ MJy sr$^{-1}$, and the highest mean PSNR, $61.405\pm1.077$ dB. BM3D gives $0.1292\pm0.0105$ and $60.952\pm0.690$, respectively. FiDeR's mean SSIM is also slightly higher. BM3D performs better for the galaxy: its RMSE and PSNR are $0.1384\pm0.0129$ and $50.873\pm0.834$, compared with FiDeR's $0.1554\pm0.0120$ and $49.860\pm0.685$. Kalman and Haar also have slightly lower galaxy RMSEs than FiDeR. Many differences among the stronger methods are comparable to the three-exposure scatter.

\begin{table*}[htbp!]
\centering
\caption{Observation-II: aperture flux changes and ellipticities in MIRI.}
\label{tab:miri_photometry}

\begingroup
\scriptsize
\setlength{\tabcolsep}{3.2pt}
\renewcommand{\arraystretch}{1.08}

\begin{tabular}{lrrrr|rrrr}
\toprule
& \multicolumn{4}{c|}{\textbf{Star: $r\leq50$ pixels}}
& \multicolumn{4}{c}{\textbf{Galaxy: $r\leq19$ pixels}} \\
\cmidrule(lr){2-5}
\cmidrule(lr){6-9}

Method
& $\Delta F_{\rm raw}$ [\%]
& $\Delta F_{\rm ref}$ [\%]
& $e$
& $\Delta e_{\rm ref}$
& $\Delta F_{\rm raw}$ [\%]
& $\Delta F_{\rm ref}$ [\%]
& $e$
& $\Delta e_{\rm ref}$ \\

&
$|\cdot|\downarrow$
&
$|\cdot|\downarrow$
&
&
$|\cdot|\downarrow$
&
$|\cdot|\downarrow$
&
$|\cdot|\downarrow$
&
&
$|\cdot|\downarrow$ \\

\midrule

Raw
& $0.000 \pm 0.000$
& $0.000 \pm 0.801$
& $0.042 \pm 0.035$
& $0.000 \pm 0.035$
& $0.000 \pm 0.000$
& $0.000 \pm 0.140$
& $0.252 \pm 0.003$
& $0.000 \pm 0.003$ \\

Mean
& $\mathbf{0.003 \pm 0.011}$
& $\mathbf{0.003 \pm 0.812}$
& $0.040 \pm 0.035$
& $-0.002 \pm 0.035$
& $\mathbf{0.003 \pm 0.017}$
& $\mathbf{0.003 \pm 0.140}$
& $0.247 \pm 0.002$
& $-0.005 \pm 0.002$ \\

Median
& $-3.380 \pm 0.149$
& $-3.380 \pm 0.871$
& $0.047 \pm 0.041$
& $0.005 \pm 0.041$
& $-1.926 \pm 0.145$
& $-1.925 \pm 0.278$
& $0.249 \pm 0.002$
& $-0.003 \pm 0.002$ \\

Kalman
& $-0.147 \pm 0.030$
& $-0.147 \pm 0.812$
& $0.042 \pm 0.037$
& $-0.001 \pm 0.037$
& $0.121 \pm 0.033$
& $0.121 \pm 0.141$
& $0.251 \pm 0.003$
& $-0.001 \pm 0.003$ \\

Haar
& $-0.013 \pm 0.023$
& $-0.012 \pm 0.823$
& $0.041 \pm 0.033$
& $-0.001 \pm 0.033$
& $-0.018 \pm 0.007$
& $-0.018 \pm 0.145$
& $0.249 \pm 0.003$
& $-0.003 \pm 0.003$ \\

NLM
& $-2.928 \pm 0.079$
& $-2.927 \pm 0.827$
& $0.034 \pm 0.014$
& $-0.009 \pm 0.014$
& $-0.166 \pm 0.018$
& $-0.166 \pm 0.143$
& $0.258 \pm 0.003$
& $0.006 \pm 0.003$ \\

BM3D
& $-0.126 \pm 0.024$
& $-0.127 \pm 0.778$
& $0.042 \pm 0.035$
& $0.000 \pm 0.035$
& $0.059 \pm 0.025$
& $0.059 \pm 0.147$
& $0.250 \pm 0.003$
& $-0.002 \pm 0.003$ \\

FiDeR
& $0.005 \pm 0.013$
& $0.005 \pm 0.814$
& $0.042 \pm 0.034$
& $\mathbf{0.000 \pm 0.034}$
& $0.039 \pm 0.021$
& $0.039 \pm 0.157$
& $0.252 \pm 0.002$
& $\mathbf{0.000 \pm 0.002}$ \\

\bottomrule
\end{tabular}

\endgroup

\par\smallskip
\parbox{0.97\linewidth}{\footnotesize
Entries are means and sample standard deviations over three exposures.
Flux differences are percentages; ellipticity differences are dimensionless.
The raw subscript denotes the same unfiltered exposure, and ref denotes the
three-exposure coadd. Ellipticity is $e=1-b/a$, measured from central second
moments within fixed apertures; it is not the intrinsic galaxy ellipticity.
The stellar aperture includes neighboring light. Arrows attached to signed
flux or ellipticity changes indicate that smaller absolute changes are
preferred; the value of ellipticity itself has no preferred numerical
direction. Boldface indicates the best value among the denoising methods;
the Raw row is treated as the unfiltered reference and is not included in
the ranking.}

\end{table*}

Table~\ref{tab:miri_photometry} complements the image metrics with signed aperture changes,
\begin{equation}
\Delta F_\mathrm{raw}=100\left(\frac{F_I}{F_\mathrm{raw}}-1\right),\quad
\Delta F_\mathrm{ref}=100\left(\frac{F_I}{F_T}-1\right),\quad
\Delta e_\mathrm{ref}=e_I-e_T.\label{eq:obsphot}
\end{equation}
We measure $e=1-b/a$ from central moments, rather than using the simulation distortion convention. FiDeR changes stellar aperture flux by $0.005\pm0.013$\% relative to the same raw exposure. Median and NLM change it by $-3.380\pm0.149$\% and $-2.928\pm0.079$\%; BM3D changes it by about $-0.126$\%. For the galaxy, FiDeR gives $0.039\pm0.021$\% and an ellipticity difference of $0.0004\pm0.0015$. The reference stellar and galaxy ellipticities are approximately 0.042074 and 0.251963. The nearly round star has substantial moment-ratio scatter, so a small signed mean shape difference does not imply a small rms error.

Only three MIRI exposures contribute to these statistics. Their standard deviations have two degrees of freedom, and self-inclusion reduces the independence of the reference. We therefore treat MIRI as a small-sample check. It supports small FiDeR aperture-flux changes and favorable stellar-aperture errors, while BM3D and Haar perform better in other regions. Differences between the NIRCam and MIRI results cannot be attributed to PSF width alone: wavelength, scene, source contrast, sampling, and calibration also differ.

\section{Conclusions}\label{sec:conclusions}
We have introduced \FiDeR, which combines finite RL reconstruction with reconvolution matched to a calibrated point-source response. The calibration includes the background-to-source ratio and approximately matches the reconstructed calibration source to the observational PSF. Under normalized, complete-domain convolution, the method also conserves the total background-subtracted flux.

The simulations show substantial suppression of pixel fluctuations with little change in bright-source widths. Faint sources retain small width offsets, so PSF preservation remains approximate and depends on the calibration regime. In the mixed stellar--galaxy field, \FiDeR\ reaches PSNRs of 46.701 and 44.850 dB at backgrounds of 10 and 20 counts per pixel, with low wing, aperture-flux, and shape errors. It outperforms the comparison denoisers in these diagnostics, while BM3D gives higher SSIM. The NIRCam observations show lower core and combined-region errors; the three-exposure MIRI test supports stellar-aperture and photometric benefits, with other methods performing better for some full-image and galaxy metrics.

\FiDeR\ achieves substantial noise reduction with little change in the measured PSF in the tested regimes, and improves on conventional denoisers in several scenes. These properties can benefit repeated exposures before UPDC deconvolution coaddition \citep{wang2022updc,wang2025updec}, machine-learning inputs, and morphological measurements. In FiDNoBa photometry \citep{zhang2026finoba}, \FiDeR\ could provide the filtered image, with its corresponding residual used for noise backfilling. The expected gains in these applications should be measured with end-to-end tests that account for the covariance of the processed pixels.

\par
\appendix
\section{Diagnostic definitions}\label{app:diagnostics}\label{sec:diagnostics}
For an output source image $I$ and reference $T$, we use the residual $d_i=I_i-T_i$. In simulations, $T$ is the noiseless PSF-convolved scene. For a region $\Omega$,
\begin{align}
\mathrm{RMSE}_\Omega&=\left[\frac{1}{|\Omega|}\sum_{i\in\Omega}d_i^2\right]^{1/2},\label{eq:rmse}\\
\mathrm{PSNR}_\Omega&=20\log_{10}\frac{\Delta T_\Omega}{\mathrm{RMSE}_\Omega},\qquad
\Delta T_\Omega=\max_\Omega T-\min_\Omega T.\label{eq:psnr}
\end{align}
We use the same reference range for every method in a comparison. Mean PSNR is the mean of realization-level logarithms. The mixed-field wing mask is the union of source-centered annuli $7\leq r\leq14$ pixels.

The structural similarity index measure (SSIM; \citealt{wang2004}) is
\begin{equation}
\mathrm{SSIM}(I,T)=
\frac{(2\mu_I\mu_T+C_1)(2\sigma_{IT}+C_2)}
{(\mu_I^2+\mu_T^2+C_1)(\sigma_I^2+\sigma_T^2+C_2)},\label{eq:ssim}
\end{equation}
with $C_1=(0.01\Delta T)^2$ and $C_2=(0.03\Delta T)^2$. We use $11\times11$ Gaussian windows with standard deviation 1.5 pixels and population covariance. Simulation scores exclude the window boundary; observational aperture scores require complete windows within the measurement region.

For fixed aperture $A_s$, $F_s(I)=\sum_{i\in A_s}I_i$ in counts. Surface-brightness images require pixel solid angles and the appropriate unit conversion. The simulated flux mean absolute error (MAE) is
\begin{equation}
\mathrm{MAE}_F=\frac{100}{N_\mathrm{src}}\sum_s
\left|\frac{F_s(I)-F_s(T)}{F_s(T)}\right|.\label{eq:photmae}
\end{equation}
The same apertures and centers are used for all realizations. They are selected from the noiseless annular signal-to-noise threshold, with maximum radius 14.5 pixels in the mixed field. The reference includes neighboring light, so this is blended-aperture rather than deblended photometry.

Simulated shapes use flux-weighted moments $Q_{ab}=\sum_{A_s}I_i u_{a,i}u_{b,i}/\sum_{A_s}I_i$ about the known source position, and
\begin{equation}
\chi=\frac{\sqrt{(Q_{xx}-Q_{yy})^2+4Q_{xy}^2}}{Q_{xx}+Q_{yy}},\qquad
\theta=\frac12\operatorname{atan2}(2Q_{xy},Q_{xx}-Q_{yy}).\label{eq:chi}
\end{equation}
Table~\ref{tab:mock2} gives mean absolute differences from the noiseless apertures. Position-angle (PA) differences are reduced modulo $180^\circ$ to $[0,90^\circ]$; profiles with $\chi<0.05$ are excluded from the PA statistic. MIRI instead uses central moments about the measured centroid and $e=1-\sqrt{\lambda_\mathrm{min}/\lambda_\mathrm{max}}$.

For stellar widths, we fit a circular Gaussian plus a constant to both output and reference, with a fixed true center and $r\leq8$ pixels. For their fitted FWHMs $w_s$ and $w_{0,s}$,
\begin{equation}
\delta_s=100\left(\frac{w_s}{w_{0,s}}-1\right),\qquad
\mathrm{scatter}=\left[\frac{\sum_{s=1}^{36}(\delta_s-\overline\delta)^2}{35}\right]^{1/2}.\label{eq:width}
\end{equation}
Figure~\ref{fig:widths} uses Monte Carlo mean images. Its scatter is across stars in percentage points, rather than across noise realizations.

\section{Conservation and calibrated response}\label{app:conservation}
For positive predicted intensities and normalized convolution on a complete periodic or otherwise flux-complete domain,
\begin{align}
\sum_i x_i^{(j+1)}
&=\sum_i x_i^{(j)}\sum_l H_{li}\frac{y_l}{(Hx^{(j)})_l}\\
&=\sum_l\frac{y_l}{(Hx^{(j)})_l}\sum_i H_{li}x_i^{(j)}=\sum_l y_l.\label{eq:rlsum}
\end{align}
Applying this identity to the calibration image and subtracting $b$ gives $\sum K_{e,k}^{(\beta)}=1$. The unit zero-frequency coefficient then gives Equation~\ref{eq:flux}. Masks, truncated support, pixel floors, and cropping can change these identities on a selected domain.

With homogeneous initialization and no fixed floors, multiplying the input by $c>0$ multiplies every RL iterate by $c$. Thus Equation~\ref{eq:effective} depends on $b/F_\mathrm{c}$. At $b=0$, the noiseless isolated-point response is independent of flux. In general, RL is nonlinear and $\RL_k(y_1+y_2)\ne\RL_k(y_1)+\RL_k(y_2)$. An isolated-point calibration therefore does not establish an identical response for every crowded or extended scene. PSF error, subpixel phase, source flux, and background error should be tested in the intended application.

\section{Background and noise treatment}\label{app:noise}\label{sec:effective_background}\label{sec:covariance}
For background-subtracted images in calibrated units, let $c$ convert intensity to effective counts and $v_\mathrm{bg}$ be the measured background variance. We use
\begin{equation}
E_\mathrm{c}=cE,\qquad b_\mathrm{eff}=c^2v_\mathrm{bg},\qquad
y_\mathrm{eff}=E_\mathrm{c}+b_\mathrm{eff}\mathbf{1},\label{eq:effectivebackground}
\end{equation}
apply Equation~\ref{eq:estimator}, and convert back by $1/c$. The restored input must be nonnegative. The adopted scale and any further offset must be recorded and included in the response calibration. This is a variance-matched approximation for read noise and resampled data, not a conversion of their full noise distribution into independent Poisson counts. A unit change alters $b_\mathrm{eff}/F_\mathrm{c}$ unless the calibration is adjusted.

Both RL reconstruction and matching convolution correlate neighboring measurements. At fixed calibration parameters, let $J_k=\partial\RL_k/\partial y$ and let $A_k$ represent convolution by $a_k$. Local linearization gives
\begin{equation}
C_\mathrm{out}\simeq A_kJ_k C_\mathrm{in}J_k^\mathsf{T}A_k^\mathsf{T}.\label{eq:cov}
\end{equation}
Independent Poisson inputs have $C_\mathrm{in}=\diag(\mu_i)$, while resampled inputs generally have off-diagonal covariance \citep{fruchter2002,makinen2020}. The variance of an aperture with weight vector $\boldsymbol w$ is $\boldsymbol w^\mathsf{T}C_\mathrm{out}\boldsymbol w$. Estimated PSF, background, and calibration flux add further uncertainty. Repeated exposures or end-to-end noise simulations can provide a practical calibration of the measurement errors.

\section{Implementation and reference conventions}\label{app:implementation}
The spatial filter labeled Kalman forms the mean $m_i$ and variance $v_i$ of the eight neighbors, clips negative roundoff variance to zero, and returns $m_i+[v_i/(v_i+\sigma_n^2)](E_i-m_i)$. It is a local adaptive filter, rather than temporal Kalman inference. Haar uses two levels and eight cyclic shifts, multiplying each detail coefficient $w$ by $1-\exp[-(|w|/(2.5\sigma_n))^6]$. NLM uses the original intensity range. BM3D scales the input and noise parameter together and returns the result to the original units. The implementation uses BM3D 4.0.3 and its BM4D dependency 4.2.5. Poisson-adapted variance stabilization and inverse transforms \citep{makitalo2011} are alternative treatments; they are not part of the baseline results reported here.

For an aperture split, the combined squared RMSE obeys
\begin{equation}
\mathrm{RMSE}_{c+w}^2=
\frac{N_c\mathrm{RMSE}_c^2+N_w\mathrm{RMSE}_w^2}{N_c+N_w}
\end{equation}
for each exposure. The displayed mean RMSEs cannot be combined this way after averaging their square roots; the combined-region values are measured directly. For a raw resampled exposure $D_j$ and equal-weight references containing $N$ exposures,
\begin{equation}
D_j-T_{-j}=\frac{N}{N-1}(D_j-T).
\end{equation}
This raw-image identity illustrates the correlation introduced by a self-including reference. Exposure standard deviations, mean errors, and source-to-source scatter are kept as separate quantities in the reported tests.

\FloatBarrier
\bibliography{references}
\bibliographystyle{aasjournal-fider}
\end{document}